\documentclass[]{aa}
\usepackage{graphicx}
\usepackage{txfonts}

\def\ms{\,m.s$^{-1}$}         

\begin{document}
\title{The $Spitzer$ search for the transits of HARPS low-mass planets - I. No transit for the super-Earth HD~40307b\thanks{The photometric time series used in this work are only available in electronic form at the CDS via anonymous ftp to  cdsarc.u-strasbg.fr (130.79.128.5) or via http://cdsweb.u-strasbg.fr/cgi-bin/qcat?J/A+A/}}
\author{M.~Gillon$^{1,2}$,  D.~Deming$^3$, B.-O.~Demory$^{4,2}$, C.~Lovis$^2$,  S.~Seager$^4$, M.~Mayor$^2$, F.~Pepe$^2$, D.~Queloz$^2$, D.~Segransan$^2$, S.~Udry$^2$, S.~Delmelle$^1$, P.~Magain$^1$}   

\offprints{michael.gillon@ulg.ac.be}
\institute{
$^1$ Institut d'Astrophysique et de G\'eophysique,  Universit\'e de Li\`ege,  All\'ee du 6 Ao\^ut 17,  Bat.  B5C, 4000 Li\`ege, Belgium \\
$^2$  Observatoire de Gen\`eve, Universit\'e de Gen\`eve, 51 Chemin des Maillettes, 1290 Sauverny, Switzerland\\
$^3$ Planetary Systems Branch, Code 693, NASA/Goddard Space Flight Center Greenbelt, MD 20771, USA\\
$^4$ Department of Earth, Atmospheric and Planetary Sciences, Department of Physics, Massachusetts Institute of Technology, 77 Massachusetts Ave., Cambridge, MA 02139, USA\\
}

\date{Received date / accepted date}
\authorrunning{M. Gillon et al.}
\titlerunning{Searching for the transit of HD~40307b with {\it Spitzer}}
\abstract{We have used {\it Spitzer} and its IRAC camera to search for the transit of the super-Earth 
HD~40307b. The transiting nature of the planet could not be firmly discarded from our first photometric monitoring
of a transit window because of the  uncertainty coming from the modeling of the photometric baseline. 
To obtain a firm result, two more transit windows were observed and a global Bayesian analysis of the three
IRAC time series and the HARPS radial velocities was performed. Unfortunately, any transit of the planet during the 
observed phase window is firmly discarded, while the probability that the planet transits but that the eclipse was 
 missed by our observations is nearly negligible (0.26\%).
 
   }

\maketitle

\section{Introduction}

Transiting  extrasolar planets are and will remain key objects for our study and understanding 
of the large planetary population hosted by our galaxy. Except for the planets of our own solar 
system, transiting exoplanets are the only ones with accurate estimates of mass, radius, and, by 
inference, constraints on internal composition. Furthermore, their special geometrical configuration gives us
the opportunity to study directly their atmospheres without the challenging need to spatially 
resolve their light from that of their host star. Occultation photometry and spectroscopy has been
used to map the thermal emission of more than a dozen highly irradiated giant planets and to 
assess their atmospheric composition, thermal gradient and albedo (see, e.g., Deming 2009). 
More demanding in terms of signal-to-noise ratio (SNR), transit transmission spectroscopy 
has been used to detect atomic and molecular features in the atmospheres of the two giant 
planets HD~209458b and HD~189733b (e.g., Charbonneau et al. 2002, Vidal-Madjar et 
al. 2003, Swain et al. 2009). These transit and occultation measurements  have opened 
a new field of astronomy: exoplanetology. 

One of the next major steps in this field will be the first atmospheric characterization of a 
$terrestrial$ extrasolar planet. Radial velocity (RV) surveys have opened the road to this 
goal by revealing a population of planets of a few Earth masses (e.g., Udry et al. 2007, Mayor et
 al. 2009a, 2009b), the so-called `super-Earths'. Last year, the CoRoT space mission detected the
  first of these super-Earths that transits its parent star: CoRoT-7b (L\'eger et al. 2009, Queloz et al. 2009). 
  With a mass $M_p = 4.8 \pm 0.8$ $M_{\oplus}$ and a radius $R_p = 1.7 \pm 0.1$ $R_{\oplus}$, CoRoT-7b 
  has an average density of $5.6 \pm 1.3$ g cm$^{-3}$, similar to the Earth's. Still, its structure is quite uncertain: its 
measured mass and radius are consistent with a rocky planet, but also with a water-rich planet that 
could contain up to 40\% of ice water (Valencia et al. 2009). Due to the relative faintness of the host 
star (V=11.7, K=9.8) and the small planet-star size ratio ($3.4 \times 10^{-4}$), a thorough characterization 
of this planet is unfortunately not possible with  existing or planned instruments. 

The detection of a second transiting super-Earth  was recently announced by the MEarth Project 
(Charbonneau et al. 2009).  This new planet orbits around the nearby M-dwarf GJ~1214 (M4.5V, 
$d = 13$ pc). Its mass is $M_p = 6.5 \pm 1$ $M_{\oplus}$,  similar to the one of CoRoT-7b. Nevertheless, 
GJ~1214b has a much larger radius than CoRoT-7b: $R_p = 2.68 \pm 0.13$ $R_{\oplus}$. The resulting 
average density  of $1.9 \pm 0.4$ g cm$^{-3}$ suggests the presence of a significant gas component. 
GJ~1214b is thus more a kind of `mini-Neptune' than a massive rocky planet, but its actual composition 
remains uncertain, as shown by Rogers \& Seager (2009). GJ~1214b is a favorable target for atmospheric 
measurements able to provide valuable insights into its composition and history. In particular,  the future 
JWST telescope (Gardner et al. 2006) will have the potential to thoroughly  characterize the atmosphere of this 
super-Earth, mainly because of the infrared brightness ($K = 8.8$) and small size (0.21 $R_\odot$) of its host star.

The future studies made possible by the detection of GJ~1214b outline how important are the few 
low-mass planets that must transit nearby bright stars. Due to their need to observe thousands of stars at 
once, most transit surveys 
do not target the brightest and most nearby stars. On the contrary, Doppler surveys target a large number 
of very bright stars. In this context, it is not surprising that the two best studied hot Jupiters, HD~209458b 
and HD~189733b, were first detected by RV measurements and caught afterwards in transit (Charbonneau et al. 
2000, Henry et al. 2000, Bouchy et al. 2005). This was also the case for the hot Neptune GJ~436b (Butler et al. 2004, 
Gillon et al. 2007b), an excellent target too for detailed follow-up atmospheric studies (see, e.g., Deming et al. 2007, 
Demory et al. 2007). Searching for the transits of the low-mass planets detected by RV measurements is 
thus an obvious method to detect transiting super-Earths suitable for a thorough atmospheric characterization.
 Doppler surveys have now  detected enough nearby low-mass planets  to make highly probable that at least 
 some of them transit their parent stars. In particular, our HARPS Doppler survey (Mayor et al. 2003) has now detected 
 more than 40 low-mass planets (Lovis et al. 2009). Among them are several announced hot Neptunes 
 (see, e.g., Santos et al. 2004, Bouchy et al. 2009)  and super-Earths (see, e.g., Udry et al. 2007, Forveille et al. 2008), 
 but also many more firmly confirmed planets that are waiting for a final analysis or some more RV measurements 
 to fully characterize their system. 

Differential photometric precisions of 0.1 \% are now routinely achieved from the ground, making rather easy the
detection of the transit of a hot Jupiter.  Due to the small size of its host star, 
the transit of the hot Neptune GJ~436b could also be detected from the ground (Gillon et al. 2007b). 
For most of the low-mass planets detected by HARPS around solar-type stars,
a transit detection requires a much more challenging photometric precision, $\sim$50 ppm or better on 
 a time scale of a few hours (corresponding to the mean transit duration for a close-in planet), and
for very bright stars. From the ground, such a precision is presently out of reach. While the photon noise would
not be a problem even for relatively small telescopes, any atmospheric instability creates a
correlated noise with a magnitude larger than the one cited above. Differential
photometry helps reducing the amplitude of this correlated noise, and ground-based sub-mmag
transit light curves nearly free of correlated noise have been obtained for a few
transiting planets (see, e.g., Gillon et al. 2009, Johnson et al. 2009, Winn et al. 2009), but for rather faint 
stars (V $>$ 10) with several comparison stars of similar brightness nearby 
(at a few arcmin at most), making possible an excellent correction of  atmospheric effects. For bright 
nearby stars, such an optimal configuration is not available. Furthermore, the transit ephemeris 
of the very low-mass planets detected by HARPS have generally a 2-$\sigma$ probability interval of 
a dozen of hours or more. This is not only because of the small amplitude of the RV signal due to the planet, but also to 
the RV low-frequency noise of the host star and the apparently large occurence of $multiple$ low-mass
 planetary systems (Lovis et al. 2009). Due to the day-night cycle and to the fact that only a fraction of some nights 
 is suitable for precise photometry of the target, ground-based telescopes are definitely not the best option for 
 such a transit search, and a space-based instrument is required. 
 
 The needed space telescope has to be 
 able to perform exquisitely precise photometry for bright stars but also to monitor continuously the 
 same star during dozens of hours. We have concluded that the best choice for such a program 
 would be the {\it Spitzer Space Telescope} (Werner et al. 2004). Due to its heliocentric orbit, it can 
 monitor most of the stars for several weeks during their visibility windows, but it has also demonstrated at many 
 instances its excellent photometric potential, detecting signals with an amplitude of  few hundreds of ppm 
 (see,e.g., Beaulieu et al. 2007). We have thus set up a {\it Spitzer} program devoted to the search of the transits 
of HARPS low-mass planets. This program is presently divided in two {\it Spitzer} sub-programs: a cycle 5 DDT
 program (ID 495) devoted to only the planet HD~40307b (Mayor et al. 2009a, hereafter M09), and a 100 hours cycle 6 DDT  program (ID 60027) devoted to $\sim$ 10 other low-mass planets.

This paper presents our results for HD~40307b.  This  super-Earth ($M_p \sin i = 4.2 M_\oplus$) 
orbits around a bright ($V = 7.2 , K =4.8$)  nearby (12.8 pc) K2-dwarf. Because of its proximity to its host star ($a = 0.047$ AU), its geometric transit probability is $\sim 7 \%$, making it a good candidate for a transit search. Section~2 presents the data used in this work, including the {\it Spitzer} observations and their 
reduction. Our data analysis is presented in Sect.~3, and our results are presented and discussed in 
Sect.~4. We give our conclusions in Sect.~5.

\section{Data}

\subsection{Radial velocities}

We used the 135 HARPS measurements presented in Mayor et al. (2009a) to estimate the HD~40307b transit windows for our {\it Spitzer} observations. We also used these data in our final global analysis aiming to assess the transit probability of the planet (see Sect.~3).
 
\subsection{{\it Spitzer} IRAC photometry} 

HD~40307b was first observed with the Infra-Red Array Camera (IRAC, Fazio et al. 2004) of {\it Spitzer} on 31 October 2008 from 01h02 UT to 09h40 UT, corresponding to a $\sim$ 2-$\sigma$ transit window of the planet as predicted by our RVs. With magnitude $K=4.8$ and K2V spectral type, HD~40307 is a very high signal-to-noise ratio (SNR) target for all IRAC bands. Considering the substantial experience of our team with {\it Spitzer} photometry at 8 $\mu$m (e.g., Deming et al. 2007, Gillon et al. 2007a, Demory et al. 2007), we conservatively decided to observe in this channel (SiAs detector) and to use the established technique of continuous staring in non-dithered subarray mode. We choose to use an exposure time of 0.32s, the largest one for which the star would not be saturated on the detector. The {\it Spitzer} flux density estimator tool\footnote{http://ssc.spitzer.caltech.edu/tools/starpet/}  predicts a flux density of 806 mJy for HD~40307b at 8 $\mu$m, which translates into a SNR of $\sim$5100 for a 1 minute integration when taking account the photon, read-out and background noises, and the instrumental throughput corrections\footnote{http://ssc.spitzer.caltech.edu/documents/som/}. 

Values for the stellar and planet sizes are needed to estimate the SNR that should be expected for the detection of
 a putative transit. Using $T_{\textrm{eff}} = 4977 \pm 59$ K, $\log{g} = 4.47 \pm 0.16$ and $[$Fe$/$H$] = -0.31 \pm 0.03$ (M09), the theoretical stellar calibration obtained by Torres et al. (2009) from well-constrained detached binary systems predicts a stellar mass of $0.78 \pm 0.06$ $M_\odot$, in good agreement with the estimation of  Sousa et al. (2008), $0.77 \pm 0.05$ $M_\odot$. The parallax of this nearby star was precisely determined by Hipparcos (ESA 1997) to $\pi = 77.95 \pm 0.53$ mas. From this value,  the Hipparcos $V$ magnitude ($V$ = 7.17, ESA 1997), the 2MASS $K$ magnitude ($K$ = 4.79, Skrutskie et al. 2006), the spectroscopic parameters from M09 and the bolometric correction calibration presented by Masana et al. (2006), we deduce a radius of $0.68 \pm 0.02$ $R_\odot$ for the star (corresponding to a luminosity $L_\ast = 0.26 \pm 0.01$ $L_\odot$).  Depending on the unknown planetary composition, the radius of HD~40307b could lie between 1.1 up to more than 5 $R_\oplus$ (Seager et al. 2007), the smallest value corresponding to an unlikely pure iron  composition. Assuming $R_\ast = 0.68 \pm 0.02$ $R_\odot$, $R_p = 1.1$ $R_\oplus$ translates into a transit depth of $220 \pm 13$ ppm and a maximum transit duration of  $\sim$135 min. Assuming an actual duration of 30 min (and thus a large impact parameter), the resulting expected SNR on the transit detection is $\sim$ 6. Thus we should have been able to detect the transit of the planet for any plausible composition and for most of the possible impact parameters. 

The data of the run are constituted of 1297 sets of 64 individual subarray images. Two of us (DD and MG) performed an independent preliminary reduction of these images.  Aperture photometry was used by both of us on the Basic Calibrated Data (BCB) provided by the standard {\it Spitzer} reduction pipeline. After a rejection of the discrepant fluxes,  the measurements were averaged for each set of 64 images, leading thus to 1297 measurements in the final light curves. The systematic effect known as the `ramp' (see, e.g., Knutson et al. 2007) is extremely sharp at the beginning of the run (see Fig.~3). In his preliminary analysis, DD simply cut off the first three hours and fitted a straight line to the remainder. Figure 1 shows the residuals of this fit $vs$ HJD. A tiny drop in brightness is visible from $\sim$2454770.70 HJD. Some tests were done to assess its significance. A transit curve having a fixed duration - calculated for a central transit - was constructed and fitted as a function of center time and transit depth. The best fit is overplotted in Fig.~1. From the precision on its depth and on the out-of-transit level, its significance is $\sim 3 \sigma$. It is thus a marginal detection. 50,000 bootstrap trials were done wherein data were permuted and the maximum amplitude of each set was found, varying the center time. Only 1-percent of the trials produced an amplitude larger than the one of the structure detected. Varying the aperture, this fraction ranged from zero to 2\%, so a 98\% confidence could be assigned to the transit-like structure under the simple line model for the baseline. 

MG cut off the first hour of data and obtained a light curve that shows a similar drop in brightness  (see Fig.~ 2). Its significance was tested with a Markov Chain Monte Carlo method (see Sect.~3). Two Markov chains were performed. For each of them, the free parameters were the transit depth, the impact parameter, the central timing and the  parameters of the baseline model (see Sect.~3 for details). The stellar mass and radius were fixed to $M_\ast = 0.78 \pm 0.06$ $M_\odot$ and $R_\ast = 0.68 \pm 0.02$ $R_\odot$. For the first chain, the initial transit depth was set to 300 ppm, while it was set to zero for the second chain.  The results for both chains were analyzed together to obtain the posterior distribution for all fitted parameters. The solutions for both chains are fully compatible, the deduced transit depth being $dF = 170^{+49}_{-62}$ ppm. The significance based on this analysis is thus comparable to the one deduced from the analysis performed by DD. The deduced transit depth translates into a planet radius $R_p  = 0.97^{+0.13}_{-0.2} R_\oplus$. Such a small radius could be estimated $a$ $priori$ unlikely, because it corresponds to a planet denser than expected for a pure iron composition. Still, any unknown systematic error in the measurements used to estimate the stellar radius could lead to a larger planet. For instance, the stellar calibration of Torres et al. (2009) leads to a larger stellar radius, $0.86 \pm 0.17$ $R_\odot$. We thus chose not to discard our marginal transit detection on the basis that it was shallower than expected, and we tried to confirm it with more observations.

\begin{figure}
\label{fig:a}
\centering                     
\includegraphics[width=9cm]{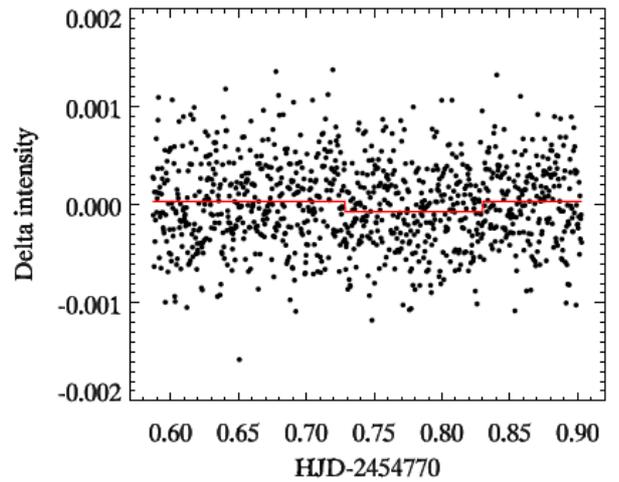}
\caption{HD~40307 {\it Spitzer} 8 $\mu$m photometry (first run) obtained after reduction by DD.
The first 3 hours have been cut off and a line was fitted in the photometry to correct for the
`ramp'. The best-fitting central transit model is overimposed. See text for details. }
\end{figure}

\begin{figure}
\label{fig:a}
\centering                     
\includegraphics[width=9cm]{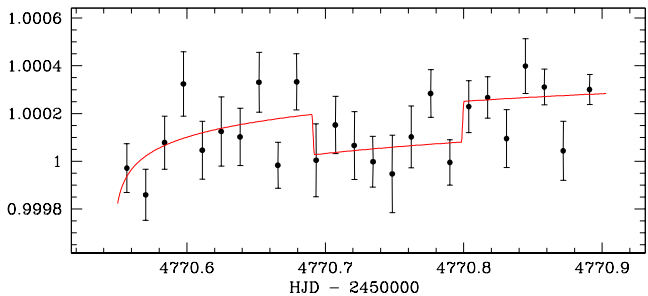}
\caption{HD~40307 {\it Spitzer} 8 $\mu$m photometry obtained after reduction by MG. The
first hour has been cut off and the points have been binned per 20 minutes. The best-fitting
ramp+transit model is overimposed. See text for details.}
\end{figure}

We requested the observation of two more transit windows with {\it Spitzer}. HD~40307 was again observed with IRAC on 02 February 2009 from 21h26 to 05h52 UT and on 13 March 2009 from 16h23 to 00h48 UT. For both runs, 1108 set of 64 individual subarray images were acquired at 8 $\mu$m, again with an exposure time of 0.32s. At this stage, it was decided  to obtain a final light curve for each run and to perform a global Bayesian analysis of these light curves in addition to the HARPS measurements to assess the probability that HD~40307b was transiting (see Sect.~3). For the three {\it Spitzer} runs, we converted fluxes from the {\it Spitzer} units of specific intensity (MJy/sr) to photon counts, and aperture photometry was obtained for HD~40307 in each image using the {\tt IRAF/DAOPHOT}\footnote{{\tt IRAF} is distributed by the National Optical Astronomy Observatory, which is operated by the Association of Universities for Research in Astronomy, Inc., under cooperative agreement with the National Science Foundation.} software (Stetson, 1987). A circular aperture with a radius of 6 pixels was centered in each image by fitting a Gaussian profile on the target. A mean sky background was measured in an annulus extending from 16 to 24 pixels from the center of the aperture, and subtracted to the measured flux for each image. For each set of 64-images, a 3-$\sigma$ median clipping was used to reject outliers, then the remaining values were averaged and the error on the mean was considered as the error on the resulting measurement. The mean number of measurements rejected per set for the three runs together was 1.5. At this stage, a 4-$\sigma$ clipping median filtering was used to reject outlier sets. For the first run, two measurements were rejected out of 1297. For both the second and third run, three measurements were rejected out of 1108. The resulting light curves binned with an interval of five minutes are shown in Fig.~3

\section{Data analysis}

The three {\it Spitzer} 8 $\mu$m  time series were first binned with an interval of five minutes to speed up the analysis. 
These binned photometric time series and the HARPS measurements were used as input data in a global analysis aiming to determine the probability that HD~40307b transits and, if so, its transit and physical parameters. The analysis was performed with the adaptative Markov Chain Monte Carlo (MCMC) algorithm presented in Gillon et al. (2010). The assumed models were based on  a star and three planets on a Keplerian orbit about their common center of mass (the planet-planet interactions were first assessed and revealed to be negligible). The planet HD~40307b had its orbital inclination free and was thus allowed to transit the star. To model the eclipse photometry, we used the photometric eclipse model of Mandel \& Agol (2002) multiplied by a systematic effect model. For the RVs, a classical Keplerian model was used in addition to a linear drift (see M09) and a Rossiter-McLaughlin (RM) effect model (Gim\'enez 2006) for the RVs obtained 
 during a transit. 
 
Considering the very red bandpass of our {\it Spitzer} photometry, we assumed no limb-darkening for the photometry. For the RVs, a quadratic limb-darkening was assumed. The values  $u_1 = 0.613$ and $u_2 = 0.161$ were deduced from ClaretÕs tables (2000; 2004) for the stellar parameters presented in M09 and for the V-filter, corresponding to the maximum of transmission of HARPS. These values were kept fixed in the MCMC. 

The 8 $\mu$m IRAC photometry is known to be affected by  a systematic effect causing the gain to increase asymptotically over time for every pixel, with an amplitude depending on their illumination history (see e.g. Knutson et al. 2008 and 
references therein). Following Charbonneau et al. (2008),  this `ramp' was modelled in each IRAC time series 
as a quadratic function of $ln(dt)$:
\begin{equation} 
A(dt)  = a_1 + a_2 ln(dt) + a_3 (ln(dt))^2 + a_4 dt \textrm{,} 
\end{equation}
where $dt$ is the elapsed time since 15 min before the start of the run. The linear term $a_4 dt$ was added to take 
into account that HD~40307 is a slowly rotating  K-dwarf star that could be variable on a time scale of several weeks, and 
that such a low-frequency variability could translate into a slight slope in the {\it Spitzer} time series.

The photometric correlated noise and the RV jitter noise were taken into account as in Gillon et al. (2010). 
For the three planets orbiting HD~40307, the jump parameters in the analysis were
the two Lagrangian parameters $e \cos{\omega}$ and $e \sin{\omega}$ 
where $e$ is the orbital eccentricity and $\omega$ is 
the argument of periastron, the orbital period $P$, the $K_2$ parameter characterizing the amplitude
 of the orbital RV signal (see Gillon et al. 2010) and the time $T_0$ for which the true anomaly 
 $M = \pi/2 - \omega$. For a transiting planet, $T_0$ represents the time of minimum light.
 For HD~40307b, the planet/star area ratio $(R_p /R_s )^2$ and the impact parameter 
 $b' = a \cos{i}/R_\ast$ were also jump parameters. 
 
Because of the expected small planet-star area ratio and the small rotational
velocity of the star ($V \sin{I} < 1$ km.s$^{-1}$, M09), the amplitude of the RM anomaly 
in case of transit of HD~40307b would be smaller than 
the typical precision of the HARPS measurements, so we kept fixed the projected 
rotational spin velocity of the star $V \sin{I}$ to 1 km.s$^{-1}$ and the projected spin-orbit angle $\beta$
of HD~40307b to 0 deg. 

A value for the stellar mass and radius were drawn at each step 
of the Markov chains from the distributions $N(0.78,0.06^2)$ $M_\odot$ and 
$N(0.68,0.02^2)$ $R_\odot$.  A uniform prior distribution was assumed for each
 jump parameter. For each {\it Spitzer} time series, the four parameters of the baseline model 
 were determined at each step of the MCMC by least-square minimization (see Gillon et al. 2010 for details).
It was also the case for the systemic RV of the star. The planet radius was forced to range from 
1 to 6 $R_\oplus$ to avoid unrealistic solutions. 

The analysis was divided into two steps. A first Markov chain was performed to assess the level 
of correlated noise in the photometry and the jitter noise in the RVs and to update the
measurement error bars accordindly. A value of 0.85 \ms  was deduced for the RV jitter noise,
while $\beta_{\rm \textrm{} red} = [1,1,1.14]$ were deduced for the photometric time series (see Gillon et al. 2010
for details). Five new chains (10$^6$ steps each) 
were then performed using the updated measurement error bars. The good convergence and mixing of these
 five chains was checked succesfully using the Gelman and Rubin (1992) statistic. 

\begin{figure}
\label{fig:a}
\centering                     
\includegraphics[width=9cm]{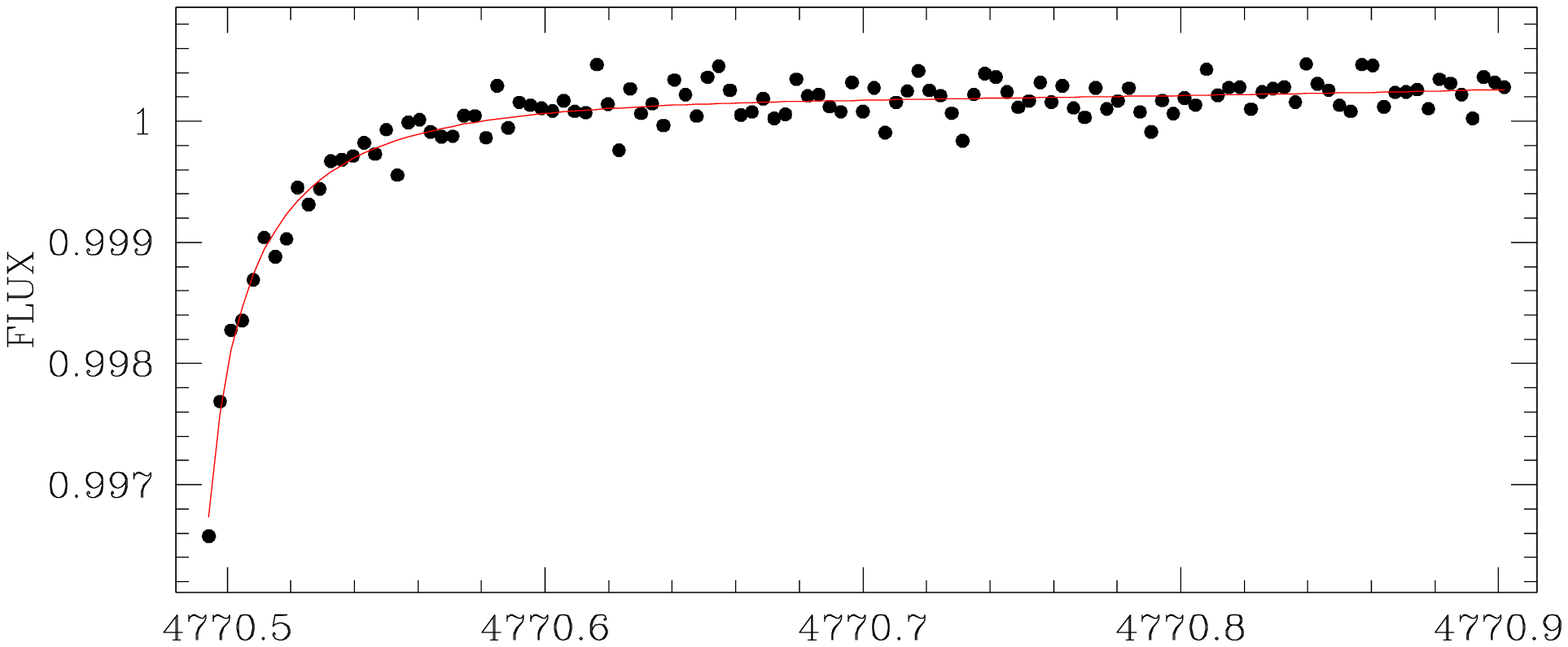}
\includegraphics[width=9cm]{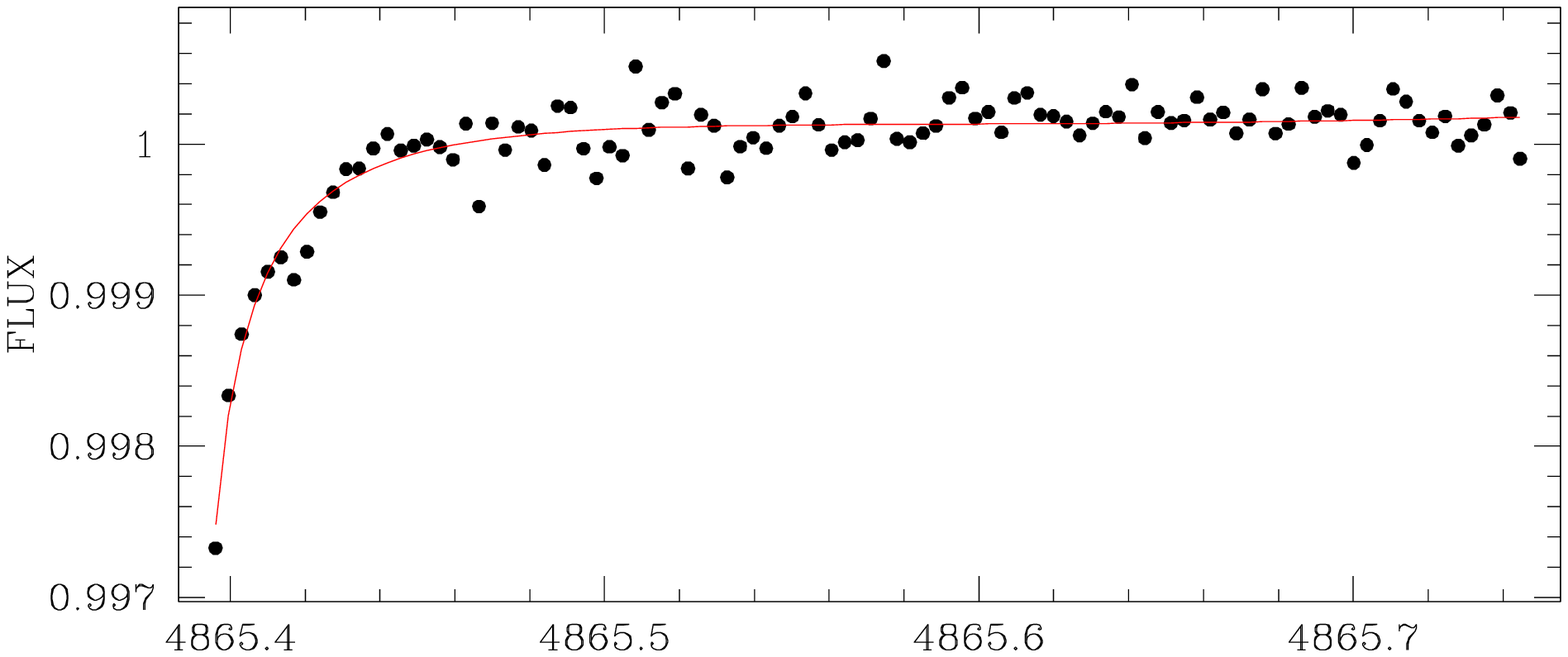}
\includegraphics[width=9cm]{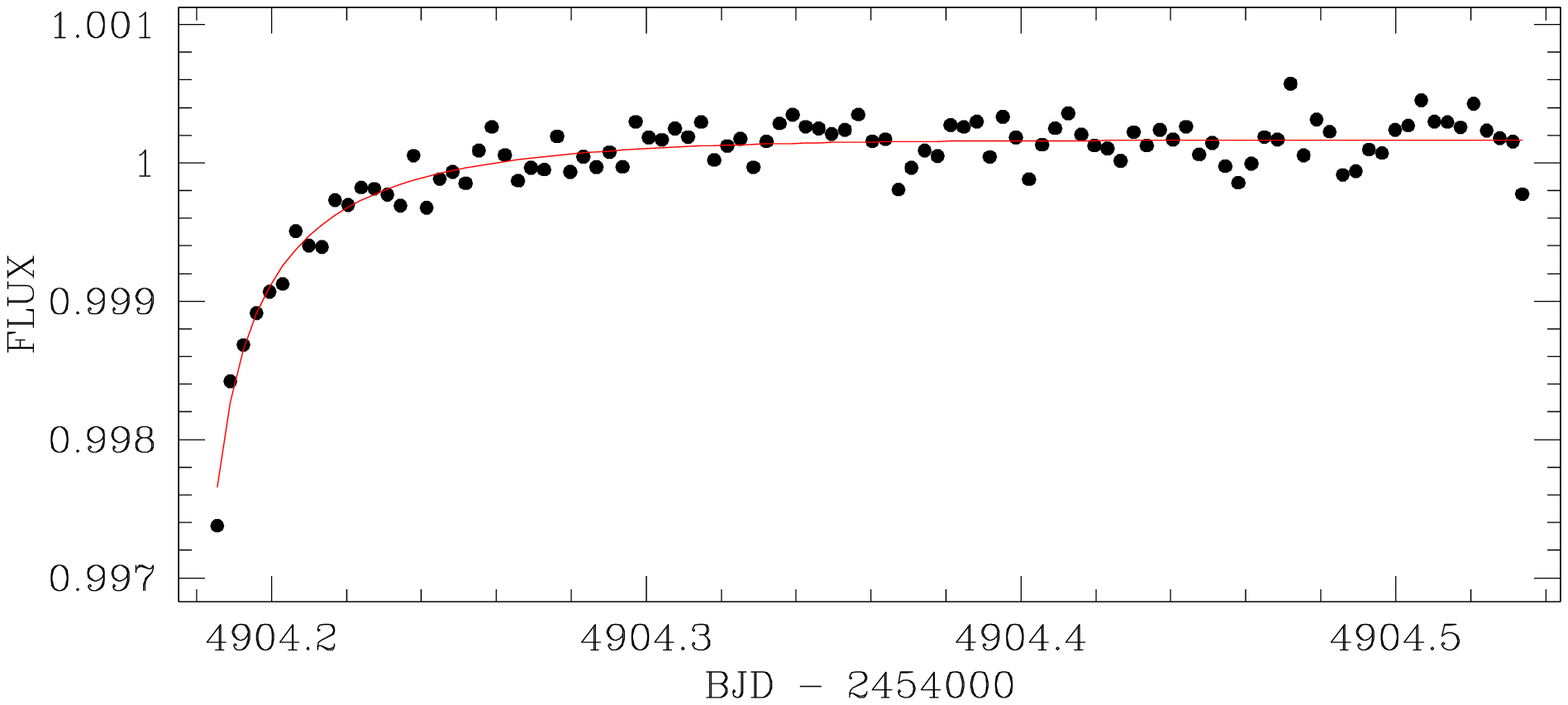}
\caption{HD~40307 {\it Spitzer} 8 $\mu$m photometry binned with an interval of five minutes.
The best-fitting model is overimposed. See text for details.}
\end{figure}

\section{Results and discussion}

Table 1 shows the median values and 68.3\% intervals of the posterior distributions
for the parameters of the three planets. These values are in good agreement with the ones reported by M09.
Figure 3 shows the three IRAC time series with the best-fitting model overimposed. No transit is present for 
the best-fitting solution. The best-fitting RV models are showed with the RV measurements in Fig.~4.

\begin{table*}
\begin{center}
\label{tab:params}
\begin{tabular}{lcccccl}
\hline
Parameter  & HD~40307b & HD~40307c & HD~40307d & Units \\ \noalign {\smallskip}
\hline \noalign {\smallskip}
Jump parameters &   & & &  \\ \noalign {\smallskip}
\hline \noalign {\smallskip}
Transit epoch  $ T_0$ - 2450000 & $4770.78 \pm 0.11$  & $4957.73 \pm 0.26$ & $4946.02^{+0.50}_{-0.51}$ &  HJD  \\ \noalign {\smallskip}
Orbital period  $ P$ & $4.31109^{+0.00051}_{-0.00052}$ & $9.6183 \pm 0.0024$ & $20.4434^{+0.0095}_{-0.0092}$ &    days  \\ \noalign {\smallskip}
$e\cos{\omega}$ & $0.009 \pm 0.060$ & $0.031^{+0.049}_{-0.048}$  &  $0.016^{+0.050}_{-0.049}$  & \\ \noalign {\smallskip}
$e\sin{\omega}$  & $0.038^{+0.057}_{-0.059}$ &  $0.027^{+0.050}_{-0.049}$ & $0.048 \pm 0.042$ & \\ \noalign {\smallskip}
RV $K_2$ & $3.22 \pm 0.19$ & $5.09 \pm 0.25$ & $7.05^{+0.34}_{-0.33}$ &    \\ \noalign {\smallskip}
Planet/star area ratio  $ (R_p/R_s)^2 $ & $0.0022^{+0.0015}_{-0.0011}$ & -  & -  & \\ \noalign {\smallskip}
$ b'=a\cos{i}/R_\ast $ &  $7.3^{+4.7}_{-4.3}$ & - & - &   $R_*$  \\ \noalign {\smallskip}
\hline \noalign {\smallskip}
Deduced parameters &   & & &  \\ \noalign {\smallskip}
\hline \noalign {\smallskip}
RV $K$ & $1.99 \pm 0.12$ & $2.40 \pm 0.12$ & $2.59 \pm 0.12$ &  \ms   \\ \noalign {\smallskip}
Orbital semi-major axis $a$ & $ 0.0477^{+0.0012}_{-0.0013}$ &$ 0.0815^{+0.0020}_{-0.0021}$& $ 0.1348^{+0.0033}_{-0.0035}$ &AU \\ \noalign {\smallskip}
Orbital eccentricity $e$ & $0.077^{+0.047}_{-0.038}$ & $0.068^{+0.042}_{-0.034}$ & $0.072^{+0.040}_{-0.034}$ & \\ \noalign {\smallskip}
Argument of periastron $\omega $ & $78^{+82}_{-73}$  & $37^{+60}_{-74}$  & $72^{+57}_{-50}$  & deg \\ \noalign {\smallskip}
$M_p \sin{i} $ & $4.27^{+0.35}_{-0.33}$ & $6.80^{+049}_{-0.52}$ & $9.28^{+0.69}_{-0.63}$ & $M_\oplus$ \\ \noalign {\smallskip}
Orbital inclination $i$ & $61^{+17}_{-24}$ & - &  - & deg \\ \noalign {\smallskip}
\hline \noalign {\smallskip}
\end{tabular}
\caption{Parameters deduced for the three planets orbiting around HD~40307 from our MCMC analysis. Jump parameters are the model parameters that are randomly perturbed at each step of the MCMC.}
\end{center}
\end{table*}

\begin{figure}
\label{fig:a}
\centering                     
\includegraphics[width=9cm]{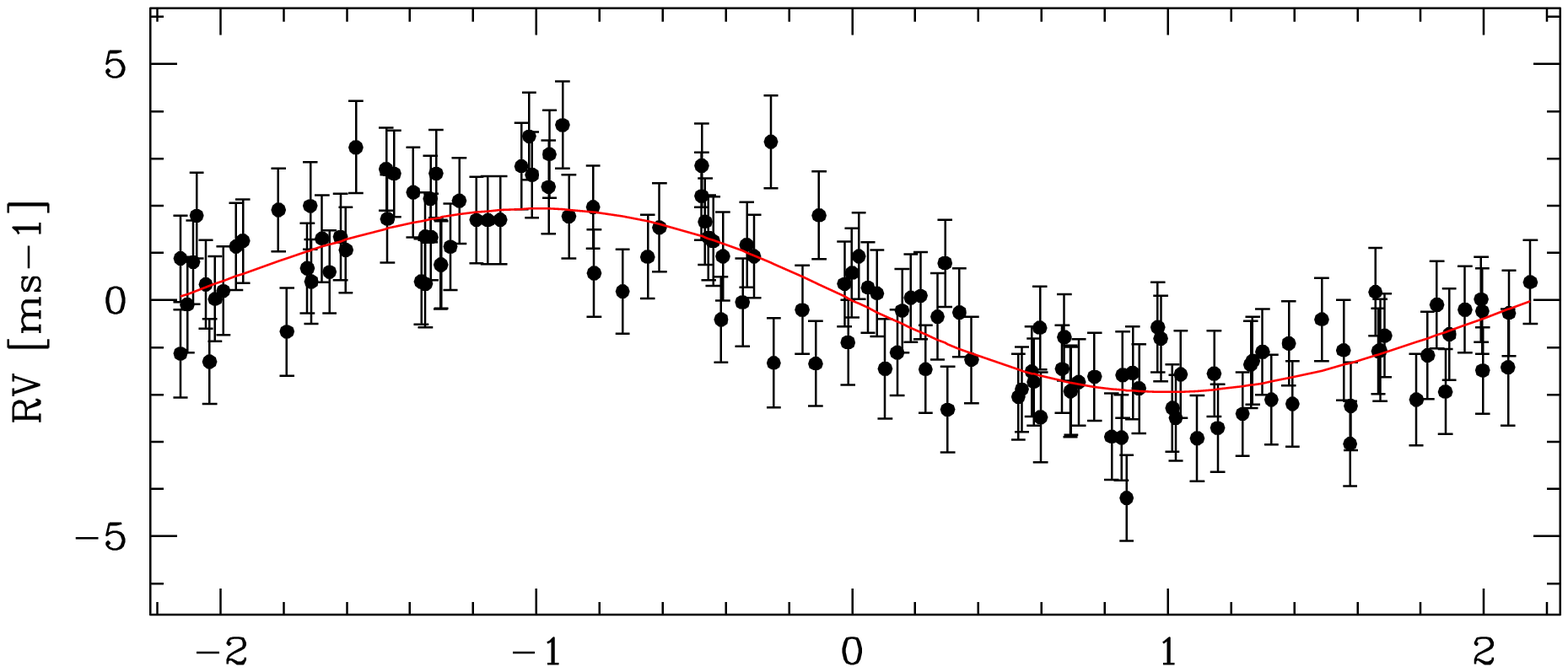}
\includegraphics[width=9cm]{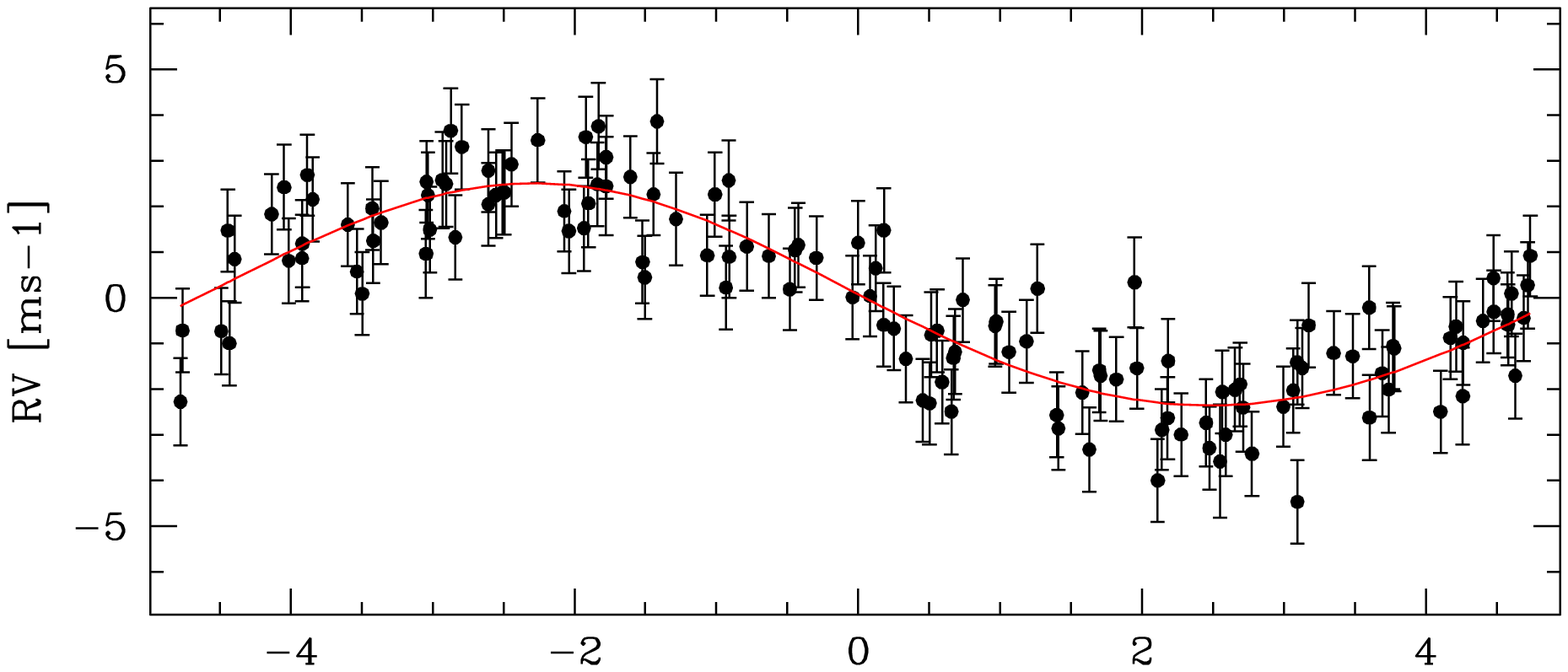}
\includegraphics[width=9cm]{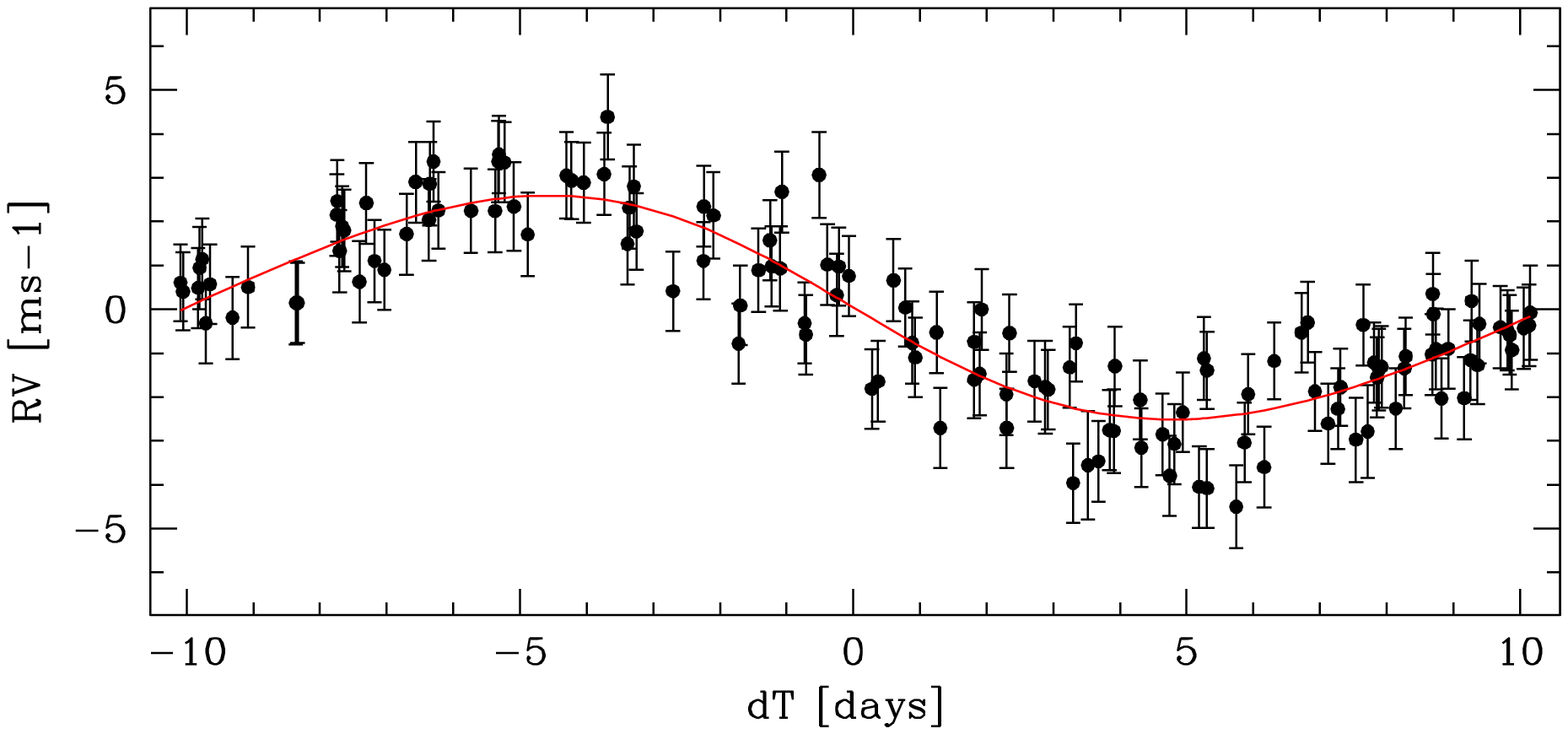}
\caption{Period-folded HARPS measurements and Keplerian curve for each of the planets, 
after correction of the effect of the two other planets and the drift. $Top$: HD~40307b, 
$middle$: HD~40307c, $bottom$: HD~40307d.}
\end{figure}

Our global Bayesian analysis leads to a posterior transit probability for HD~40307b of 0.3\%. 
If we consider only full transits, the posterior probability is only 0.19\%. The
marginal transit detection that we obtained from the first IRAC run is thus unfortunately not confirmed, 
and the transiting nature of HD~40307b is rejected at $\sim 3\sigma$.
As shown in Fig.~5, the fraction of the simulations for which HD~40307b transits concern
mostly  transits that happen {\it outside} the observed window. The resulting
posterior probability that a transit was present in the observation window is 0.04\%, 
while no simulation resulted in a full transit present in the IRAC runs.
The presence of a transit in the {\it Spitzer} data is thus rejected with a high level of
 confidence, while the probability that the planet transits but that the eclipse was 
 missed is nearly negligible (0.26\%). 

\begin{figure}
\label{fig:a}
\centering                     
\includegraphics[width=9cm]{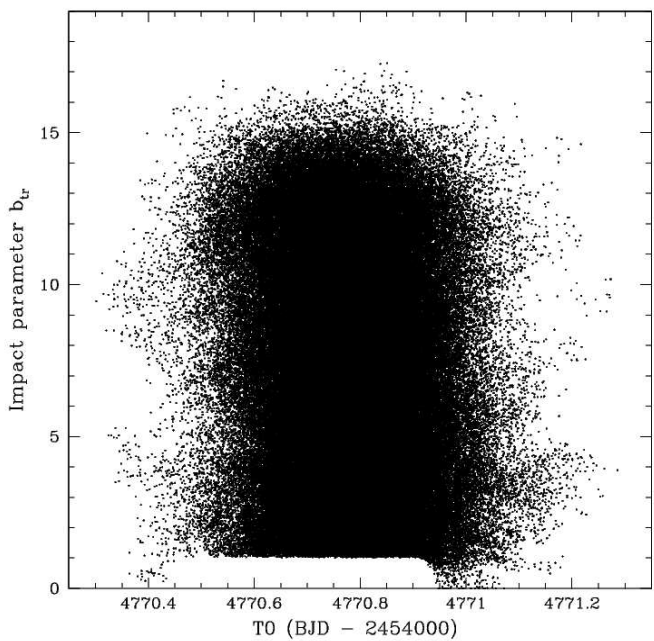}
\caption{Posterior probability density: time of minimum light $T_0$ $vs$ transit impact parameter $b_{tr}$. }
\end{figure}

To investigate the effect of  the stellar size on this result, we performed a second MCMC 
analysis assuming $R_\ast = 0.86 \pm 0.17$ $R_\odot $ instead of $0.68 \pm 0.02$ $R_\odot$ (see 
discussion in Sect.~2). The obtained transit probability was 0.63 \%, leading to the conclusion that 
a transit of HD~40307b can be considered as very unlikely, even if the star is actually significantly
larger than deduced from its parallax and bolometric luminosity.

Figure 6 shows the IRAC photometry corrected for the baselines of the three time series and period-folded
on the best-fitting $T_0$. The $rms$ of this light curve is 143 ppm for a mean interval of 1.9 minutes, while the $rms$
goes down to 29 ppm when the light curve is binned with an interval of 30 minutes. This illustrates 
well the high photometric potential that the cryogenic  {\it Spitzer} had at 8 $\mu$m. Now that
its cryogen is depleted,  only the 3.6 and 4.5 $\mu$m channels remain operational, and similar 
photometric precisions are not guaranteed, mainly because of the inhomogeneous intra-pixel sensitivity of 
the InSb detectors of IRAC (e.g. Knutson et al. 2008 and references therein).

\begin{figure}
\label{fig:a}
\centering                     
\includegraphics[width=9cm]{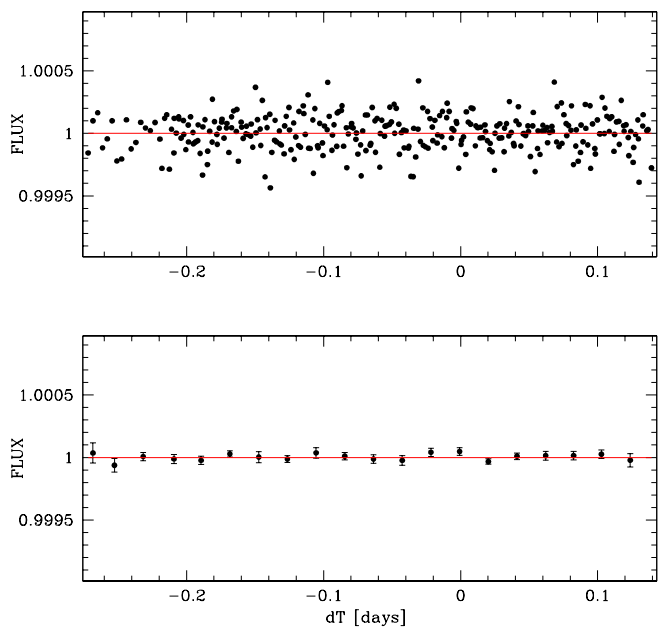}
\caption{$Top$: Light curve obtained after correction for the systematics of the three IRAC light curves used 
in this work and folding on the best-fitting ephemeris. $Bottom$: same but after binning with an interval
of 30 minutes. }
\end{figure}

If present, the transit of a planet as small as 1.1 $R_\oplus$ (220 ppm) should have been easily detected. 
To check this, we injected a fake transit of  $\sim$1 hour and  220 ppm in the {\it Spitzer} photometry, 
using the values reported in Table 1 for $T_0$ and $P$. Our first MCMC analysis of these data was composed of
two chains of 10$^6$ steps. Except in the `burn-in' phases of the chains, a transit was present in all the 
steps of both chains, indicating a firm detection. To estimate a formal transit probability, we performed two 
new chains for which no transit was allowed. For both models (with and without transit), the marginal likelihood was
computed from the MCMC samples like described by Chib \& Jeliaskov (2001). The deduced Bayes ratio (e.g., Carlin \& Louis 2008) 
was $1.4\times10^{23}$ in favor of the model with a transit, confirming the very firm transit detection. Fig. 7 shows
 the obtained best-fitting transit model superimposed on the period-folded IRAC photometry corrected for the systematics. 
The deduced transit depth is $232 \pm 23$ ppm, in good agreement with the actual value of 220 ppm.

\begin{figure}
\label{fig:a}
\centering                     
\includegraphics[width=9cm]{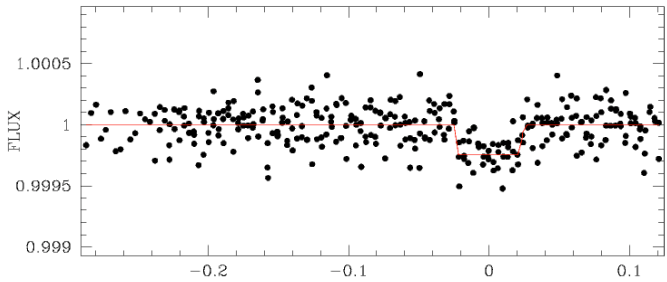}
\caption{Period-folded photometry with the best-fitting transit model obtained after insertion of a fake transit 
with a depth of 220 ppm and a duration of 1 hour. }
\end{figure}

We investigated  the origin of the `transit' marginally detected in the first IRAC time series alone. It appears that it 
was most probably due to the baseline model used in our preliminary analysis. To recall, DD fitted a line in the 
light curve after cutting off the first three hours, while MG cut off the first hour and assumed a quadratic 
logarithmic function of time for the ramp model, i.e. the same function than shown in Eq. 1 but without the 
$a_4 dt$ term. Figure 8 shows the photometry obtained after correction by the best-fitting baseline model obtained
by least-square minimization, using Eq.~1 for the model, with and without the $a_4 dt$ term. The residual 
correlated noise visible in the time series obtained when the $a_4 dt$ term is not used could be easily `compensated'
by a shallow transit, while the correlated noise is nearly negligible when the $a_4 dt$ term is used. 
Using only the first IRAC light curve and the HARPS measurements as data, we performed two new 
MCMC analysis, one with and one without the $a_4 dt$ term in the baseline equation. The stellar 
radius was set to 0.86 $R_\odot$. Without the $a_4 dt$ term, the deduced transit probability was 97.1\%, 
while it dropped to 0.9\% with it. Comparing both models, the resulting Bayes ratio was 7.4 in favor of the model
with the $a_4 dt$ term. We  conclude thus that the significance of our transit detection
in the first IRAC light curve was in fact smaller than estimated in our preliminary analysis. 
This illustrates well how careful one should be when searching for very shallow transits in photometric time 
series presenting systematic effects of much larger amplitude, even if these systematics are rather well 
characterized. 

\begin{figure}
\label{fig:a}
\centering                     
\includegraphics[width=9cm]{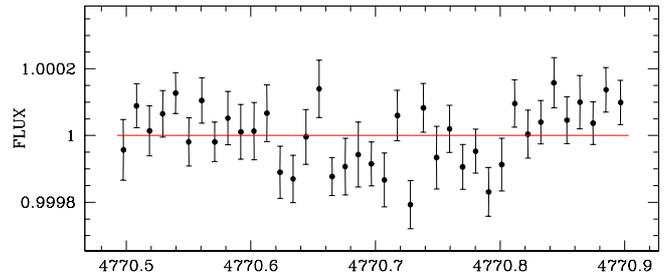}
\includegraphics[width=9cm]{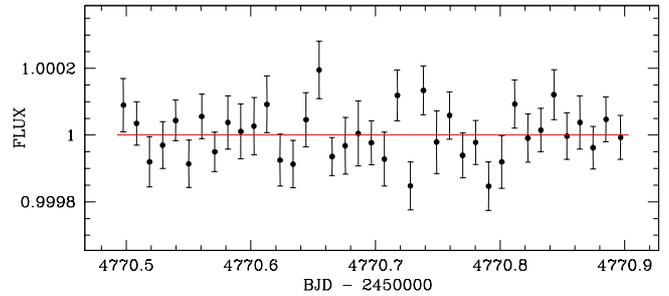}
\caption{Light curves for the first run corrected for the baseline model and binned with a five minutes
interval, without ($top$) and with ($bottom$) the linear term $a_4 dt$ in Eq.~1. }
\end{figure}

\section{Conclusions}

We have used {\it Spitzer} and its IRAC camera to search for the transit of the super-Earth 
HD~40307b. The transiting nature of the planet could not be firmly discarded from our first photometric monitoring
of a transit window because of the  uncertainty coming from the modeling of the photometric baseline. 
To obtain a firm result, two more transit windows were observed and a global Bayesian analysis of the three
IRAC time series and the HARPS radial velocities was performed. Unfortunately, any transit of the planet during the 
observed phase window is firmly discarded, while the probability that the planet transits but that the eclipse was 
 missed by our observations is nearly negligible (0.26\%).
 
 \begin{acknowledgements} 
This work is based in part on observations made with the {\it Spitzer Space Telescope}, which is operated by the Jet Propulsion Laboratory, California Institute of Technology under a contract with NASA. Support for this work was provided by NASA. M. Gillon acknowledges support from the Belgian Science Policy Office in the form of a Return Grant, and thanks J. Montalb\'an for her valuable suggestions on bolometric corrections. 

\end{acknowledgements} 

\bibliographystyle{aa}

\end{document}